\documentclass[a4paper,twoside,twocolumn]{article}
\usepackage[T1]{fontenc} 
\usepackage[utf8]{inputenc}
\usepackage[english]{babel}

\usepackage{authblk}
\usepackage{csquotes}
\usepackage{amssymb}
\usepackage{amsmath}
\usepackage{siunitx}
\usepackage{graphicx}
\usepackage{dcolumn}
\usepackage{bm}
\usepackage{xcolor}
\usepackage{physics}
\usepackage{xfrac}
\usepackage{booktabs}

\usepackage{blindtext}
\usepackage{hyperref}
\hypersetup{
    unicode=false,          
    pdftoolbar=true,        
    pdfmenubar=true,        
    pdffitwindow=false,     
    pdfstartview={FitH},    
    pdfauthor={Kevin Multani},     
    colorlinks=true,       
    linkcolor=blue,          
    citecolor=blue        
}

\usepackage[bbgreekl]{mathbbol}

\DeclareMathAlphabet{\mathbbold}{U}{bbold}{m}{n}
\newcommand{\Am}{A}
\newcommand{\Av}{A}
\newcommand{\Wv}{w}
\newcommand{\pel}{\ell}
\newcommand{\fre}{t}

\newcommand{\lst}{L}
\definecolor{RED}{HTML}{FF0000}

\newcommand{\ul}{_{\text{\textsc{ul}}}}

\usepackage[hmarginratio=1:1,top=32mm,columnsep=18pt,%
left=0.78in,right=0.78in]{geometry} 
\usepackage[small,labelfont=bf,up,textfont=it,up]{caption}

\usepackage{abstract} 

\usepackage{titlesec} 
\titleformat{\section}[block]{%
\large\bfseries}{\thesection.}{1em}{}
\titleformat{\subsection}[block]{\bfseries}{\thesubsection.}{1em}{} 

\usepackage[sorting=none,citestyle=numeric-comp, natbib]{biblatex}
\bibliography{biblio}

\title{\Large\bfseries Sequential hypothesis testing for Axion Haloscopes}

\author[1]{Andrea \textsc{Gallo Rosso}\thanks{\href{mailto:andrea.gallo.rosso@fysik.su.se}
{andrea.gallo.rosso@fysik.su.se}.}}
\author[2]{Sara \textsc{Algeri}}
\author[1]{Jan \textsc{Conrad}}
\affil[1]{Physics Department and Oskar Klein Centre, 
Stockholm University, Stockholm, Sweden}
\affil[2]{School of Statistics, University of Minnesota, %
Minneapolis, MN, USA}

\date{}

\begin{document}

\twocolumn[
    \maketitle
    \begin{@twocolumnfalse}
    \vspace{-1.5em}
    \begin{abstract}
        The goal of this paper is to introduce a  novel likelihood-based
        inferential framework for axion haloscopes which is valid under
        the commonly applied ``rescanning'' protocol.  The proposed method
        enjoys  short data acquisition times and a simple tuning of the
        detector configuration.  Local statistical significance and power
        are computed analytically, avoiding the need of burdensome 
        simulations. Adequate corrections for the look-elsewhere effect
        are also discussed. The performance of our inferential strategy
        is compared with that of a simple method which exploits the
        geometric probability of rescan. Finally, we exemplify the
        method with an application to a HAYSTAC type axion haloscope.
    \end{abstract}
    \vspace{2.5em}
    \end{@twocolumnfalse}
    ]
\saythanks


\section{Introduction}

The existence of dark matter was first postulated almost a century
ago \citet{Zwicky:1937zza} and is today a well-established
paradigm in astrophysics and cosmology. However, its detection and
identification on a fundamental level is still one of the most
investigated problems across all physical sciences
\cite{Bergstrom:2000pn,%
Bertone:2004pz,Jaeckel:2010ni,Feng:2010gw,%
Profumo:2019ujg,Arbey:2021gdg,Green:2022hhj}.
Among the plethora of viable particle candidates which may
constitute dark matter, axions are some among the most appealing ones.
In fact, they were originally theorized in relation to the strong
CP problem of fundamental physics \cite{Peccei:1977hh,%
Peccei:1977ur,Weinberg:1977ma,Wilczek:1977pj}. Only later,
physicists  realized  axions exhibit properties that are
consistent with those of cold dark matter and their
behavior is compatible with cosmological and astrophysical
constraints \cite{Preskill:1982cy,Abbott:1982af,Dine:1982ah}
--- see  
\cite{DiLuzio:2020wdo,Choi:2020rgn,Chadha-Day:2021szb} for recent reviews.

At present, the experimental effort towards a direct detection
is witnessing many experiments being planned or already running
\cite{Graham:2015ouw,Irastorza:2018dyq,Billard:2021uyg,%
Irastorza:2021tdu,Semertzidis:2021rxs,Adams:2022pbo}.
One of the most well-established way to search for axions is
through cavity microwave experiments, generally referred to as
haloscopes. The idea was first proposed by Sikivie in 1985
\cite{Sikivie:1985yu}; it relies on the Primakoff effect
\cite{Primakoff:1951iae} to induce the axion-to-photon conversion
in a resonant apparatus by matching the Compton wavelength of
the axion and the resonant mode of the cavity. The interaction
appears as power being deposited at a frequency matching the
axion mass, i.e.\ $\nu_a = m_a c^2 h^{-1}$.

Since the axion frequency $\nu_a$ is unknown \emph{a priori},
a standard axion experiment is conceptually run by sequentially
tuning the detector over the available frequencies. The goal is
that of identifying a peak emerging from the fluctuations in the
noise power of the system. The ratio between the power deposited
by the axion over the noise power determines the signal-to-noise
ratio, which is directly linked to the sensitivity and improves
with the time invested measuring.
As a result, the extent to which an axion experiment is able to
explore the parameter space does not depend just on the detector
properties and the available resources, such as the total
lifetime, but also on the way those resources are 
effectively allocated. In fact, while experimenters could
(in principle) focus solely on one specific frequency and increase
the signal-to-noise ratio arbitrarily,  they have also
the freedom to carefully evaluate the fluctuations measured over
multiple frequencies and repeat the measurement only for those 
those deemed to provide promising hints of an axion.
It follows that the choice of such protocol for both data
acquisition and statistical analyses is crucial. 

Despite a variety
of protocols have been proposed 
\citep[e.g.,][]{Foster:2017hbq,Knirck:2018knd,Palken:2020wgs,%
Ahn:2020nyv}, the main source of disagreement across different
experiments is how to flag plausible candidate frequencies to be
rescanned. This especially true when aiming to perform an analysis
in a frequentist framework. 

In this manuscript, we aim to address this inconsistency by
introducing a statistically rigorous, likelihood-based, rescan
protocol which provides the tools needed for a direct comparison
of the results of future experiments, such as the reachable
sensitivity, or upper limits on the axion-photon coupling. We
believe that this is  especially needed given that
there is no general consensus on the way such
quantities are defined within the community.

The proposed inferential framework 
is presented in Sections \ref{sec:likelihood}-\ref{sec:LEE}.
In order to ease the exposition, the main elements of the
procedure are introduced in a step-by-step manner, gradually increasing
level of complexity. Specifically, in Section \ref{sec:likelihood},
we outline the frequentist likelihood-based approach for axion
searches  in the simple  scenario where only one scan is performed
at a given frequency.
In Section \ref{sec:rescan},  we introduce the re-scan protocol,
i.e., we allow our measurements to be conducted sequentially,
over multiple scans at a fixed frequency. This is the main
methodological result of the article. In Section \ref{sec:LEE},
we discuss adequate   ``look-elsewhere effect'' (LEE) corrections.
That is, we allow the re-scan protocol to be performed over multiple
frequencies. Hence, adequate LEE adjustments are introduced in
order to control the probability of a false discovery over the
entire frequency range considered.
Section \ref{sec:ulhay} outlines how to set upper limits in the context
of a real axion experiment. Specifically, we use the parameterized
properties of the first phase of HAYSTAC detector 
\cite{Brubaker:2016ktl,Brubaker:2017ohw} to make a projection
of the reachable upper limit on the photon-axion coupling
constant, showing how different definitions can affect the
final result. A summary and concluding remarks are presented in Section \ref{sec:conclusion}.

\section{Likelihood-based inference}
\label{sec:likelihood}

In what follows, we assume that the sensitivity domain of
a given axion experiment is discretized into $M$ bins, each one of them marking a frequency $\nu_i$, idexed by the subscript $i=1,\,\dots,\,M$. The measuring
process associates to the $i$-th bin a set of $y_{ij}$ normalized
noise fluctuations, possibly
taken under different detector configurations $j=1,\dots,\,N_i$. We assume the fluctuations $y_{ij}$ to
be normally distributed with unknown mean, $\mu_{ij}$, and known
standard deviation, $\sigma_{ij}$, i.e.,\ 
\begin{equation}\label{eq:ypdf}
    y_{ij} \sim \mathcal{N}(\mu_{ij},\,\sigma_{ij}).
\end{equation}
The normality assumption in \eqref{eq:ypdf} simplifies the calculations and is a
reasonable approximation of the true distribution of the noise fluctuations \citep[e.g.,][]{ADMX:2001dbg,ADMX:2020hay,%
Brubaker:2017rna}. In Equation \eqref{eq:ypdf},
the mean $\mu_{ij}$ is nonzero only if an axion  deposits  
power  proportional to some coupling, $\Av$, on the $i$-th frequency
bin, i.e.,
\begin{equation}\label{eq:muij}
\mu_{ij} = \Av\,\Wv_{ij}.
\end{equation}
Here, $\Av$ is the parameter of interest and the weights
$\Wv_{ij}$ are assumed to be known --- see for instance\ Equation
\eqref{eq:defwh} in Section \ref{sec:ulhay} or Ref.\ 
\cite{Millar:2022peq}.
The main goal of the proposed analysis is that of testing the hypotheses
\begin{equation}
    \label{eqn:test}
    H_0: A=0\quad\text{versus}\quad H_1: A\neq 0;
\end{equation}
that is, we aim 
 to assess if
the \emph{null hypothesis}, $H_0$, of no axion  is consistent
with the observations collected by the detector.

Denote with
$Y_{\fre}=\{y_{i1},\dots,y_{iN_i}\}_{i\in\mathcal{I}_\fre}$ the sets
of fluctuations relevant to check for the presence of an axion
at a frequency $\nu_{\fre}$, with $\fre=1,\,\dots,\,T$, and
$T\leq M$. Specifically, $Y_\fre$ collects all the  normalized noise fluctuations measured across all the $N_i$ detector configurations at frequencies $\nu_i$ in a neighborhood $\mathcal{I}_\fre$ of $\nu_t$. From \eqref{eq:ypdf} it follows that 
  the likelihood function is
\begin{equation}\label{eq:LikDef}
    \mathcal{L}_\fre(A) = \prod_{i\in\mathcal{I}_\fre}\prod_{j=1}^{N_i}
    \phi(y_{ij};\,\Am w_{ij},\sigma_{ij}),
\end{equation}
where $\phi(\:\cdot\:;\,\Am w_{ij},\,\sigma_{ij})$ denotes the
density of a normal with mean $\Am w_{ij}$ and standard deviation
$\sigma_{ij}$.

We begin by investigating the case where the presence of an axion is tested at a given frequency $\nu_\fre$. In order to simplify the notation, the subscript $\fre$ is dropped in the remainder of this section and in Section \ref{sec:rescan}. It will be reintroduced in Section \ref{sec:LEE}, when discussing the situation where multiple tests conducted simultaneously over different frequencies.

The statistic we rely upon to  test \eqref{eqn:test} is the 
\emph{signed-root likelihood ratio test} \citep[e.g.,][]{jensen}. 
When testing the hypotheses in \eqref{eqn:test}, it specifies as
\begin{equation}
    \label{eq:defSk}
    s = \frac{x}{\sqrt{u}},
\end{equation}
where
\begin{equation}\label{eq:defUkXk}
    x = \sum_{i\in\mathcal{I}}\sum_{j=1}^{N_i}
    \frac{w_{ij}y_{ij}}{\sigma_{ij}^2}
    \quad\text{and}\quad
    u = \sum_{i\in\mathcal{I}}\sum_{j=1}^{N_i}
    \frac{w_{ij}^2}{\sigma_{ij}^2}.
\end{equation}
Here, $u$ summarizes the various detector configurations,
with no need to further transform and rescale
the original data $y_{ij}$, which are now aliased by
the (one-dimensional) random variable $x$. Being a
linear combination of normally distributed random variables, $x$ is also
Gaussian with variance $\sigma_{x}^2 = u$ and mean
$\mu_{x} = A u$. The latter is valid assuming
the most favorable case, i.e.\ the hypothesis of an axion lying on the frequency being tested.
In fact, we may expect that, on a given frequency bin, an axion with (true) frequency $\nu_a$ and coupling
constant $\Av$ deposits the
same power of an axion with (slightly shifted)
frequency $\nu_{a}^{\prime}$
and coupling constant $\Av^\prime = \Av\Wv_{ij}^{\prime}/w_{ij}$.

Under $H_0$ in \eqref{eqn:test}, the distribution of the test statistic $s$ in \eqref{eq:defSk} is that of a
standard normal. Whereas, if $H_0$ is false, then a  mean shift  is introduced.
Once an experimental value $\hat{s}$
is measured, the significance of the departure from the null hypothesis of zero mean can be quantified by computing the p-value
\begin{equation}\label{eq:Pval}
    P(s \geq\hat{s}\,|\,H_0) = 1 - \Phi\left(\hat{s}\right),
\end{equation}
where $\Phi\left(\cdot\right)$ is the cumulative distribution
function of a standard Gaussian. When the probability \eqref{eq:Pval}
is below a pre-determined significance, typically denoted by $\alpha$,
then $H_0$ is rejected and a discovery is claimed.
Conventionally, the line is drawn at $5\sigma$ significance which
corresponds to
$\alpha =1-\Phi(5)= \num{2.87e-7}$. In the following, we will
refer to this as the \emph{statistical discovery} condition, neglecting any additional
procedure such as the ``manual interrogation'' of persistent
candidates \cite{Brubaker:2017rna,ADMX:2020hay}.

\section{The rescan protocol}
\label{sec:rescan}
We now extend the experimental freedom by introducing the option
to flag promising noise fluctuations as potential candidates, and
increase the data sample with subsequent measurements. Additional measurements are done sequentially, until the
fluctuation disappears or a discovery is claimed. 
Here, we will consider two main methods, the so-called
\emph{geometric test} (and adequate generalizations),
and our \emph{likelihood-based rescan protcol}. As it will become
clear in Sections \ref{sec:geometric}-\ref{sec:likelihood_rescan},
the former is essentially equivalent to the frequentist framework
discussed in \cite{Palken:2020wgs}, whereas the latter is new, and
allows us to include additional information into the analysis. 

\subsection{The geometric approach}
\label{sec:geometric}
\begin{figure}[t]
\includegraphics[width=\columnwidth]{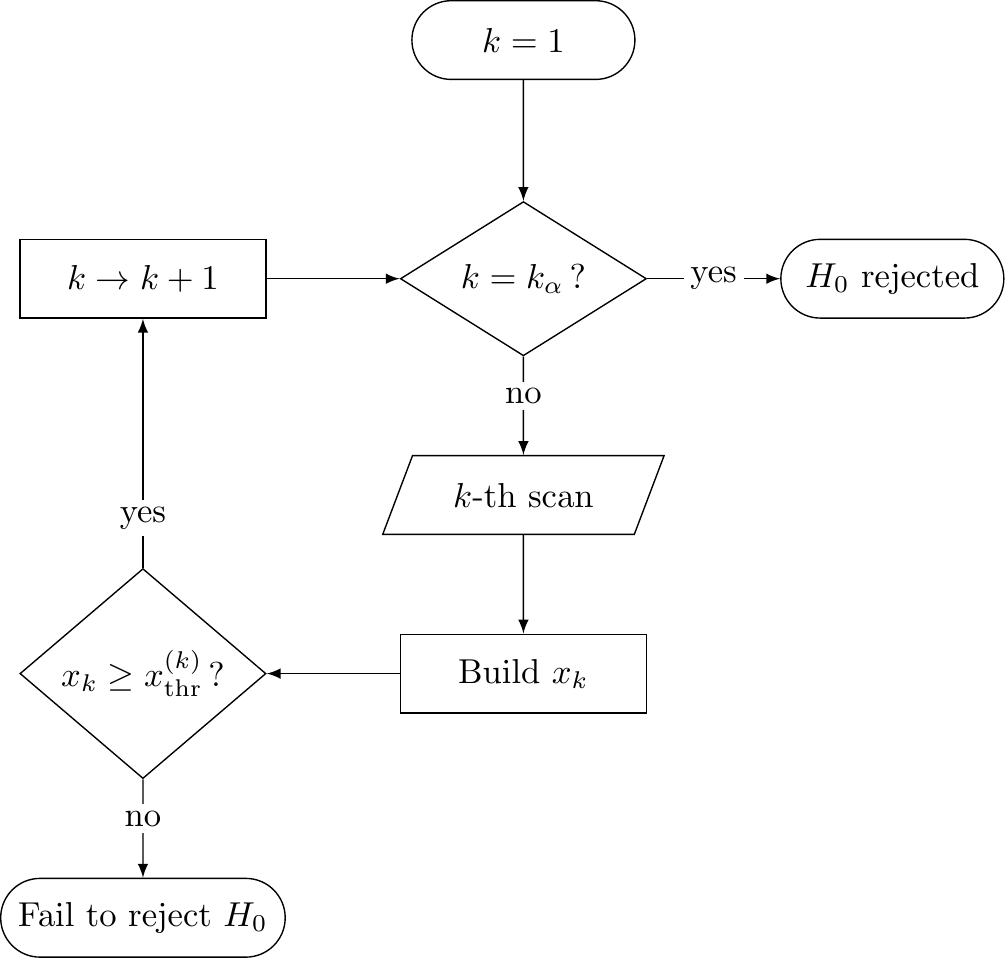}
\caption{Flowchart of the rescan protocol based on the
geometric test (Section \protect\ref{sec:geometric}).
Its outcomes are
either success (or fail) in rejecting $H_0$
and claiming (or not) the axion discovery.}
\label{fig:digG}
\end{figure}

A rather simple procedure to test for the presence of an axion
is sketched in Figure \ref{fig:digG}. The inputs required are the desired significance level,
$\alpha$, needed to claim a discovery, and the probability of 
performing an additional scan.

Specifically, denote with $y_{ij\pel}$ the fluctuation measured on bin $i$, under the $j$-th detector configuration during the $\pel$-th session of
measurement, or rescan. Let 
$x_\pel$ be the value of $x$  in \eqref{eq:defUkXk} observed at the $\pel$-th scan, i.e.,
\begin{equation}
    \label{eq:xl}
    x_\pel = \sum_{i\in\mathcal{I}}\sum_{j=1}^{N_i}\frac{w_{ij\pel}y_{ij\pel}}{\sigma_{ij\pel}^2}.
\end{equation}
Similarly to Section 
\ref{sec:likelihood}, we assume that the $x_\pel$ are independent
and normally distributed with mean depending on the coupling
constant $\Av$, i.e., $\mu_{x_\pel} = A u_{\pel}$, and variance 
\begin{equation}
    \label{eq:ul}
    u_\pel=\sum_{i\in\mathcal{I}}\sum_{j=1}^{N_i}
    \frac{w^2_{ij\pel}}{\sigma^2_{ij\pel}}.
\end{equation}

The $(\pel+1)$-th scan is performed if  the condition
\begin{equation}
\label{eq:rule_thr}
x_\pel>x_{\text{thr}}
\end{equation}
is verified for some pre-determined threshold $x_{\text{thr}}$. The latter can be chosen, for example, to ensure that the probability of rescan when no axion is present (under $H_0$) is $5\%$, i.e.,  
\begin{equation}
\label{eq:prob_thr}
P(x_\pel>x_{\text{thr}}|H_0)=0.05.
\end{equation}

We then proceed by considering the total number of scans being
performed, $K$, as our test statistic. Specifically,
\eqref{eq:rule_thr} implies that  additional
 measurements are collected until such conditions no longer holds.
Hence, $K$ can be treated as a geometric random variable with
probability of ``success'' given by the complement of  
\eqref{eq:prob_thr}. Notice that, in this context a ``success'' corresponds to the even of stopping the procedure. We can then reject the null hypothesis of
``no axion'' whenever the rescan protocol based on
\eqref{eq:rule_thr} leads to $k$ scans, with $k$ satisfying
\begin{equation}
    \label{eq:condition}
    P(K\geq k|H_0)=[P(x_\pel>x_{\text{thr}}|H_0)]^{(k-1)}\leq
    \alpha.
\end{equation}
Notice that $K$ differs from $k$ in that the former is a random
variable taking values $k\in\{1,\dots,\infty\}$. Therefore, the
number $k$ of rescans reached in a given experiment is simply the
value of $K$ observed.    

The geometric test above can be further extended by allowing the
threshold $x_{\text{thr}}$ to vary at each scan. This effectively
reduces to the frequentist approach discussed in Ref.\ 
\cite{Palken:2020wgs} and for which \eqref{eq:condition}
generalizes to
\begin{equation}
    \label{eq:condition2}
    P(K\geq k)=\prod_{\pel=1}^{k-1}
    \left[P(x_\pel>x^{(\pel)}_{\text{thr}}|H_0)\right]\leq\alpha
\end{equation}
with $x^{(\pel)}_{\text{thr}}$ denoting the pre-determined threshold
for the $\pel$-th scan.

It is worth emphasizing that 
\eqref{eq:condition}-\eqref{eq:condition2} allow us to determine
the minimum number of scans, hereafter denoted by $k_\alpha$, needed to claim a discovery
at the prescribed significance level, $\alpha$,  and for a given
set of threshold(s) $x^{(\pel)}_{\text{thr}}$. Alternatively, when
the resources available strongly limit the maximum number of
measurements that can be performed, one can tune the threshold(s) $x^{(\pel)}_{\text{thr}}$ to ensure that \eqref{eq:condition2} 
is verified for a pre-determined value of $k_{\alpha}$.

\subsection{The likelihood-based rescan protcol}
\label{sec:likelihood_rescan}
Despite the simplicity of the approach outlined in Section
\ref{sec:geometric}, its main limitation is that it requires us
to perform at least $k_{\alpha}-1$ measurements in order to
claim a statistical discovery. Moreover, as noted in \cite{Palken:2020wgs},
a design based on binary outcomes like \eqref{eq:rule_thr} is
blind to the actual magnitude of the fluctuations, leading to
a waste of potentially valuable information. 
Here, we outline a novel procedure which allows us to overcome
these limitations. Additional advantages in terms of power and
number of rescans needed to claim a statistical discovery are discussed in
Section \ref{sec:comparison}.

\begin{figure}[t]
\includegraphics[width=\columnwidth]{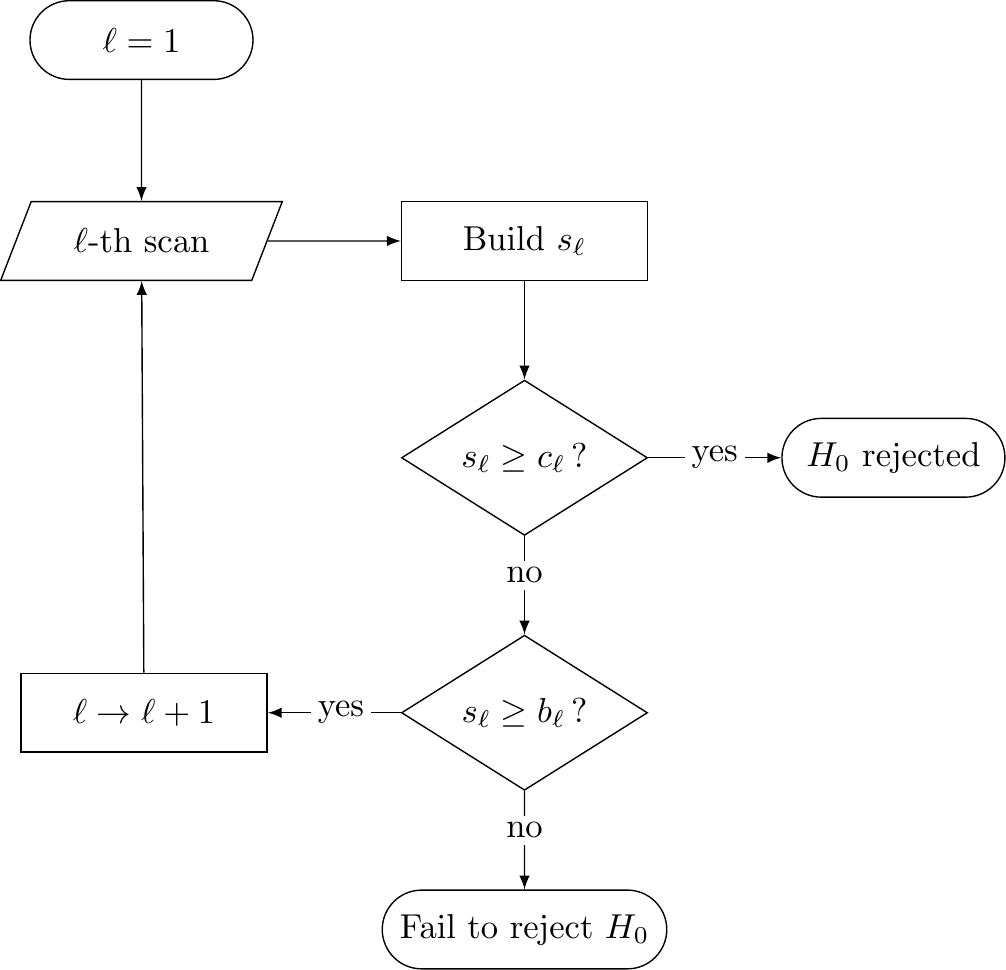}
\caption{Flowchart of the likelihood-based rescan protocol
discussed in Section \protect\ref{sec:likelihood_rescan}.
The possible outcomes are the same as the
geometric test (Figure \protect\ref{fig:digG}).
}
\label{fig:digL}
\end{figure}

The main steps of our proposed procedure are summarized in
Figure \ref{fig:digL}. We refer to this approach as
\emph{likelihood-based rescan protcol} and it consists
of the following.

Let $k_\alpha$ be the maximum number of scan to be performed.
For example, as noted at the end of Section \ref{sec:geometric},
$k_\alpha$ can be chosen on the basis of the resources available,
or to ensure a significance level of $\alpha$ using the
geometric approach.
Recall that $Y_h$ denotes the set
of fluctuations
relevant to check for the presence of an axion at a frequency $\nu_{\fre}$ in
the $h$-th scan. At the $\pel$-th scan, $\pel=1,\dots,k_\alpha$, we compute the
test statistic $s_\pel$ considering the  set of
measurements, 
$\bigcup_{h=1}^\pel Y_h$.
The updated test statistic $s_\pel$ generalizes $s$ in \eqref{eq:defSk} in that 
\begin{equation}
 \label{eq:sl}
 s_\pel=\frac{\sum_{h=1}^\pel x_h}{\sqrt{\sum_{h=1}^\pel u_h}},
\end{equation}
with $x_h$ and $u_h$ calculated as in \eqref{eq:xl} and
\eqref{eq:ul}, respectively, for each of the $h=1,\dots,\pel$ scans. We then rely on two conditions
to determine whether or not a statistical discovery should be claimed,
or if an additional scan should be performed. 

Specifically, for a suitably chosen set of constants
$\{c_\pel\}_{\pel=1}^{k_\alpha}$, once the $\pel$-th scan has been completed,
we first assess if the condition
\begin{equation}
    \label{eq:Cl}
    C_\pel = \{s_\pel\geq c_\pel\}
\end{equation}
is verified. If it is, a discovery claim is made. Whereas,
if $C_\pel$ does not hold, we verify the validity of the condition
\begin{equation}
    \label{eq:Bl}
    B_\pel = \{s_\pel\geq b_\pel\},
\end{equation}
where $b_\pel$ is the $\pel$-th elements of a set of constants
$\{b_\pel\}_{\pel=1}^{k_\alpha}$, used to control the probability
of rescan. If $B_\pel$ in \eqref{eq:Bl} holds, we proceed with
another rescan; if $B_\pel$ is not verified, the experiment is stopped,
and we declare no discovery. 
Note that in \eqref{eq:Cl} and \eqref{eq:Bl} we are implicitly
setting $b_\pel \leq c_\pel$, with equality holding for
$\pel=k_\alpha$ in order to have a definite outcome. 
It follows that, in our likelihood-based protocol, condition
\eqref{eq:Bl} plays the same role as \eqref{eq:rule_thr} in the
geometric approach. Conversely from the latter, however, by
combining $B_\pel$ and $C_\pel$, we allow researchers to claim a
statistical discovery at any step, determining a potential break in the procedure even before reaching the $k_\alpha$-th scan. 

Clearly, the statistical validity of the proposed strategy depends entirely
on the specification of the constants $b_\pel$ and $c_\pel$.
These must be chosen to ensure that the probability of a false
discovery over multiple scans is no lower than the pre-determined
significance level $\alpha$. This can be done as follows.  

Denote  with $D_\pel$ the event of claiming a statistical discovery at the $\pel$-th
scan. On the basis of the scheme in Figure \ref{fig:digL},
such event specifies as
\begin{equation}
    D_\pel = \left[\bigcap_{h=1}^{\pel-1}
    \left\{C_{h}^c\cap B_{h}\right\}\right]\cap C_\pel,
\end{equation}
with $C_{h}^c=\{s_h<c_h\}$. To ensure that the probability of a false discovery is indeed equal to the desired level  $\alpha$, we must guarantee that
\begin{equation}\label{eq:PUDn}
    P\left(\bigcup_{\pel = 1}^{k_\alpha} D_\pel\middle|\,H_0\right)
    = \sum_{\pel = 1}^{k_\alpha} P\left(D_\pel\middle|\,H_0\right)
    =\alpha,
\end{equation}
where the first equality follows from the
fact that the events $D_\pel$ are mutually exclusive,
i.e.\ their intersection is empty. Therefore,
it is sufficient to choose $b_\pel$ and $c_\pel$ such that
\begin{equation}\label{eq:PDaN}
    P(D_\pel|\,H_0)=\frac{\alpha}{k_\alpha}
\end{equation}
for all $\pel=1,\dots,k_\alpha$ to ensure that \eqref{eq:PUDn} holds.

Interestingly, condition \eqref{eq:PDaN} can easily be
expressed in closed form and thus one can exploit the latter
to derive the constants $b_\pel$ and $c_\pel$ sequentially.

Specifically, for $\pel=1$, the setup is the same as that
outlined in Section \ref{sec:likelihood}. From \eqref{eq:Pval},
it follows that
\begin{equation}
    P(D_1|\,H_0)= P(s_1 \geq c_1|\,H_0) =
    1-\Phi(c_1)=\frac{\alpha}{k_\alpha},
\end{equation}
and thus, $c_1$ is simply the quantile of order $1-\alpha$
of a standard normal. The threshold $b_1$, can be determined
by solving the equation
\begin{equation}
    \int_{b_1}^{c_1}\dd{s_1} \mathcal{N}
    \left(s_1\,\middle|\,0, 1\right) = q_1,
\end{equation}
where $q_1$ is the probability of engaging in the second scan
under the background-only hypothesis. For example, similarly to
\eqref{eq:prob_thr}, we may chose $q_1=0.05$.

Let us extend the reasoning to the less trivial case of $\pel>1$.
Since the constants $b_\pel$ and $c_\pel$ are computed sequentially,
it follows that, at the $\pel$-th scan, $b_h$ and $c_h$ are known
for all $h\leq \pel-1$. Define $\bm{x} = (x_1,\,x_2,\,\dots,x_\pel)$
to be the vector of powers $x_h$ computed as in \eqref{eq:xl}
for each of the $\pel$ scans. Similarly, denote with 
$\bm{u}=(u_1,\,u_2,\,\dots,u_\pel)$ the vector collecting the
respective configurations $u_h$ in \eqref{eq:ul}. Analogously,
we can define the vector of test statistics
$\bm{s}=(s_1,\,s_2,\,\dots,s_\pel)$, each computed as in \eqref{eq:sl}.
We can then easily derive the joint probability density function (pdf)
of the random vector $\bm{s}$ starting from  that of $\bm{x}$,
and which follows a multivariate normal of dimension $\pel$, i.e.,
\begin{equation}
\label{eq:multiN}
    \bm{x} \sim\mathcal{N}_\pel\left(\bm{\mu_x},\,\bm{\Sigma_x}\right)
\end{equation}
with
\begin{equation}
    \bm{\mu_x} = \Av\bm{u}
    \quad\text{and}\quad
    \bm{\Sigma_x} = \text{diag}(\bm{u}).
\end{equation} 

Denote with $\bm{J}$ the Jacobian of the transformation in \eqref{eq:sl} that allows us to link the vectors $\bm{x}$ and $\bm{s}$. From \eqref{eq:multiN}, it follows that
\begin{equation}
\label{eq:svec}
    \bm{s} \sim\mathcal{N}_\pel\left(\bm{s}|\bm{\mu_s}=\bm{J\mu_x}
    ,\,\bm{\Sigma_s}=\bm{J\Sigma_xJ^{\mathrm{T}}}\right).
\end{equation}
Combining \eqref{eq:svec} and \eqref{eq:PDaN} we have
\begin{align}
    &\frac{\alpha}{k_\alpha}=P(D_\pel|\,H_0)=\nonumber\\
    &P(s_\pel \geq c_\pel,b_{\pel-1}\leq s_{\pel-1}< c_{\pel-1},
    \dots,\,b_{1}\leq s_{1}< c_{1})=\nonumber\\
    &\int_{c_\pel}^\infty\dd s_\pel\int_{b_{\pel-1}}^{c_{\pel-1}}\dd
    s_{\pel-1}\dots
    \int_{b_1}^{c_1}\dd s_1\,\mathcal{N}_\pel(\bm{s}|\,\bm{0},\,
    \bm{\Sigma_s}).\label{eq:FindCn}
\end{align}
Ultimately, $c_\pel$ is obtained by solving the equation
in \eqref{eq:FindCn}.

Whereas, to determine the threshold $b_\pel$, we consider
the conditional probability of the event
$\{b_\pel\leq s_\pel<c_\pel\}$ \emph{given} the outcome of
the previous $\pel-1$ scans (with no discovery break). We require
such probability to be equal to the desired probability of rescan
$q_\pel$ (e.g., $q_\pel=0.05$, for all $\pel=1,\dots,k_\alpha$), i.e.,
\begin{align}
    &P\left(C_{\pel}^c\cap B_{\pel}\middle|\bigcap_{h=1}^{\pel-1}
    \{C_{h}^c\cap B_{h}\},H_0\right)=\nonumber\\
    &P\left(\bigcap_{h=1}^{\pel}
    \left\{C_{h}^c\cap B_{h}\right\}\middle|H_0\middle)\middle/
    P\middle(\bigcap_{h=1}^{\pel-1}
    \left\{C_{h}^c\cap B_{h}\right\}\middle|H_0\right)=\nonumber\\
    &\frac{\int_{b_{\pel}}^{c_{\pel}}\dd s_{\pel}\dots
    \int_{b_1}^{c_1}\dd s_1\,\mathcal{N}_\pel(\bm{s}|\bm{\mu_s},\,
    \bm{\Sigma_s})}{\int_{b_{\pel-1}}^{c_{\pel-1}}\dd s_{\pel-1}\dots
    \int_{b_1}^{c_1}\dd s_1\,\mathcal{N}_{\pel-1}(\bm{s}|\bm{\mu_s},\,
    \bm{\Sigma_s})}=q_\pel,\label{eq:FindBn}
\end{align}
where the first equality follows from Bayes's theorem.
Notice that in \eqref{eq:FindBn} the only unknown quantity 
is the constant $b_\pel$. 

Equation \eqref{eq:FindBn} allows a step-by-step
determination of the first $k_\alpha-1$ constants $b_\pel$.
The last one, $b_{k_\alpha}$, is chosen to be equal to $c_{k_\alpha}$
in order ensure that a conclusion is reached in no more than
$k_\alpha$ scans.

\subsection{Likelihood-based vs.\ geometric approach}
\label{sec:comparison}

\begin{table}[b]
\caption{\label{tab:table1}%
Coefficients $b_\pel$ and $c_\pel$ satisfying  \protect\eqref{eq:FindCn} and
\protect\eqref{eq:FindBn} for the toy example
described in Section \protect\ref{sec:comparison}.}
\begin{tabular}{lrrrrr}
\toprule
$\pel$ &\multicolumn{1}{c}{1} &\multicolumn{1}{c}{2}
    &\multicolumn{1}{c}{3} &\multicolumn{1}{c}{4}
    &\multicolumn{1}{c}{5}\\
\midrule
$b_\pel$ & 1.9929 & 3.1810 & 4.0916 & 4.7943 & 4.8031\\
$c_\pel$ & 5.3018 & 5.2967 & 5.2681 & 5.1820 & 4.8031\\
\bottomrule
\end{tabular}
\end{table}

\begin{figure}[t]
\includegraphics[width=\columnwidth]{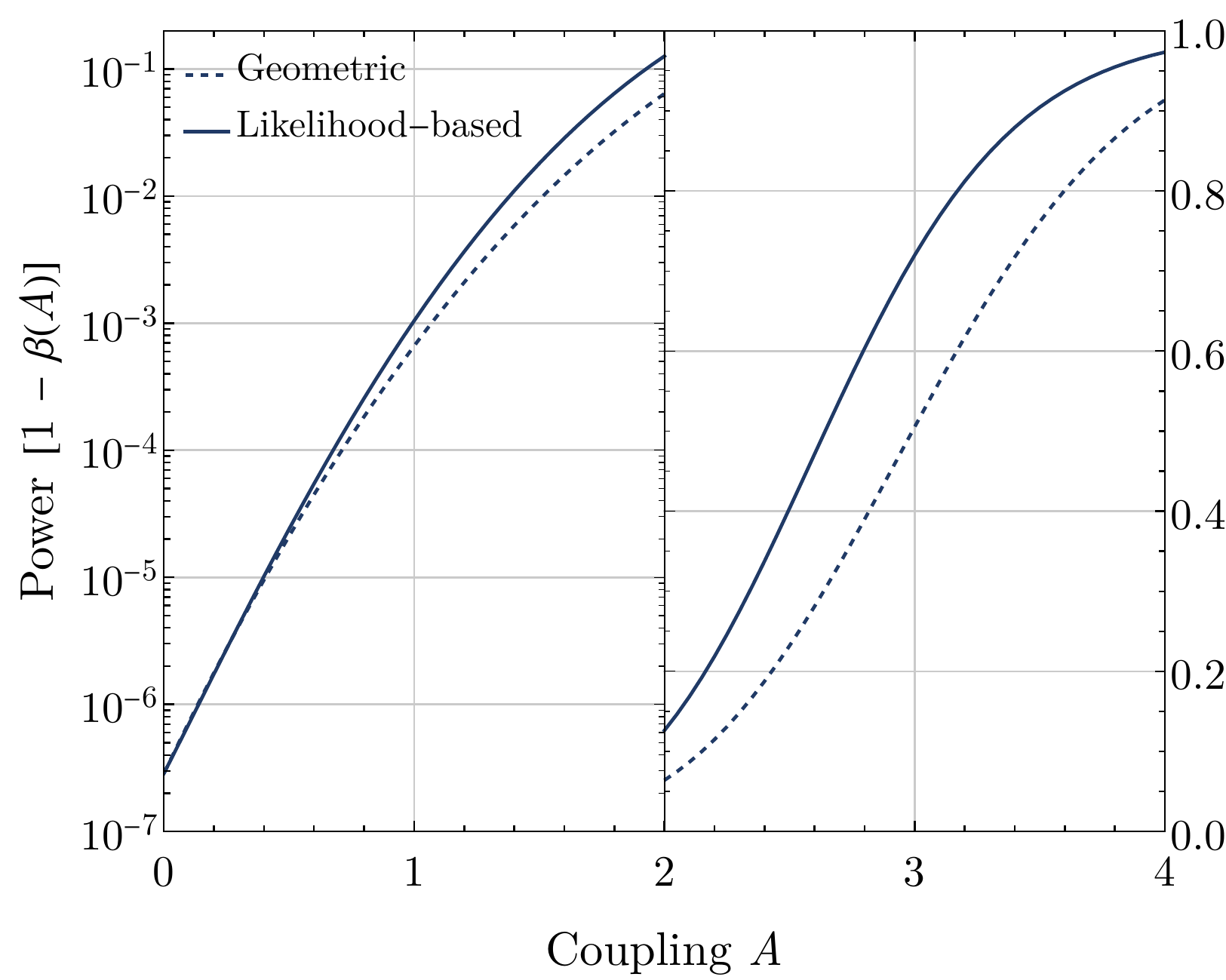}
\caption{Power as a function of the coupling $A$.
Dashed line: power for the geometric approach computed as in Equation \protect\eqref{eq:powGeo}.
Solid line: power for the likelihood-based approach computed as in Equation \protect\eqref{eq:powLik}. The values
in the range $A\in[0, 2]$ are magnified in logarithmic scale.}
\label{fig:power}
\end{figure}

In order to acquire a deeper understanding of the   advantages
and disadvantages of these two approaches, we investigate their
statistical properties by means of a toy example. Let the maximum
number of scans to be perform be $k_\alpha = 5$ and consider the
conventional $5\sigma$ level, that is, $\alpha = \num{2.8665e-7}$.
For each $\pel=1,\,\dots,\,k_\alpha$, we assume that $u_\pel=1$ and we let
\begin{equation}
    \label{eq:q1expl}
    P(x_\pel>x_{\text{thr}}|H_0)=\alpha^{1/4}.
\end{equation}
This would ensure that
the probability of false discovery is exactly $\alpha$, allowing a 
fair comparison of the power (see below and Figure \ref{fig:power}).
From condition \eqref{eq:condition}, it follows that
the threshold value, $x_{\mathrm{thr}}$, for the geometric
test is given by
\begin{equation}
    x_{\mathrm{thr}} = \Phi^{-1}(1 - \alpha^{1/4}) \approx 1.9929.
\end{equation}
The $b_\pel$ and $c_\pel$ constants required by the
likelihood-based rescan protocol have been computed as described
in Section \ref{sec:likelihood_rescan}, with 
 rescan probability as in \eqref{eq:q1expl}. The resulting values of
$b_\pel$ and $c_\pel$ are reported in Table \ref{tab:table1}.

We proceed by comparing the power of both approaches;
that is, the probability to correctly rejecting the null
hypothesis,
$H_0$ in \eqref{eqn:test}, when the alternative hypothesis, $H_1$ is true. In other words, the power corresponds to the sensitivity of our test in  detecting an axion expressed in probabilistic terms.  In statistical
literature, it is conventionally  denoted as $1-\beta$,
were $\beta$ corresponds to the probability of type II error,
i.e., the probability of failing to reject $H_0$ when $H_1$ is true. In our case, the power is a function of
the coupling $\Av$ constant and, for both the geometric and the
likelihood-based rescan approach, it can be easily computed
analytically. Specifically, for the geometric test, since
$u_\pel = 1$, for all $\pel=1,\dots,k_\alpha$, we have
\begin{equation}
 \label{eq:powGeo}
    1 - \beta(\Av) = \left[1- \Phi(x_{\mathrm{thr}}
    - \Av)\right]^{k_\alpha - 1};
\end{equation}
whereas, the power of the likelihood-based rescan protocol is 
\begin{equation}
    \label{eq:powLik}
    1 - \beta(\Av) = 
    \sum_{h = 1}^{k_\alpha} P(D_h|\,\Av).
\end{equation}
In \eqref{eq:powLik}, the probabilities $P(D_h|\,\Av)$ can be calculated similarly to $P(D_h|\,H_0)$ in \eqref{eq:FindCn}.
The power functions in  \eqref{eq:powGeo}
and \eqref{eq:powLik} are plotted in  Figure \ref{fig:power}.

As expected, since the likelihood-based protocol uses more
information, it also exhibits higher power;
especially for larger values of the coupling constant, $\Av$.
The two methods perform similarly  when  $\Av$ approaches zero.

\begin{figure}[t]
\includegraphics[width=0.96\columnwidth]{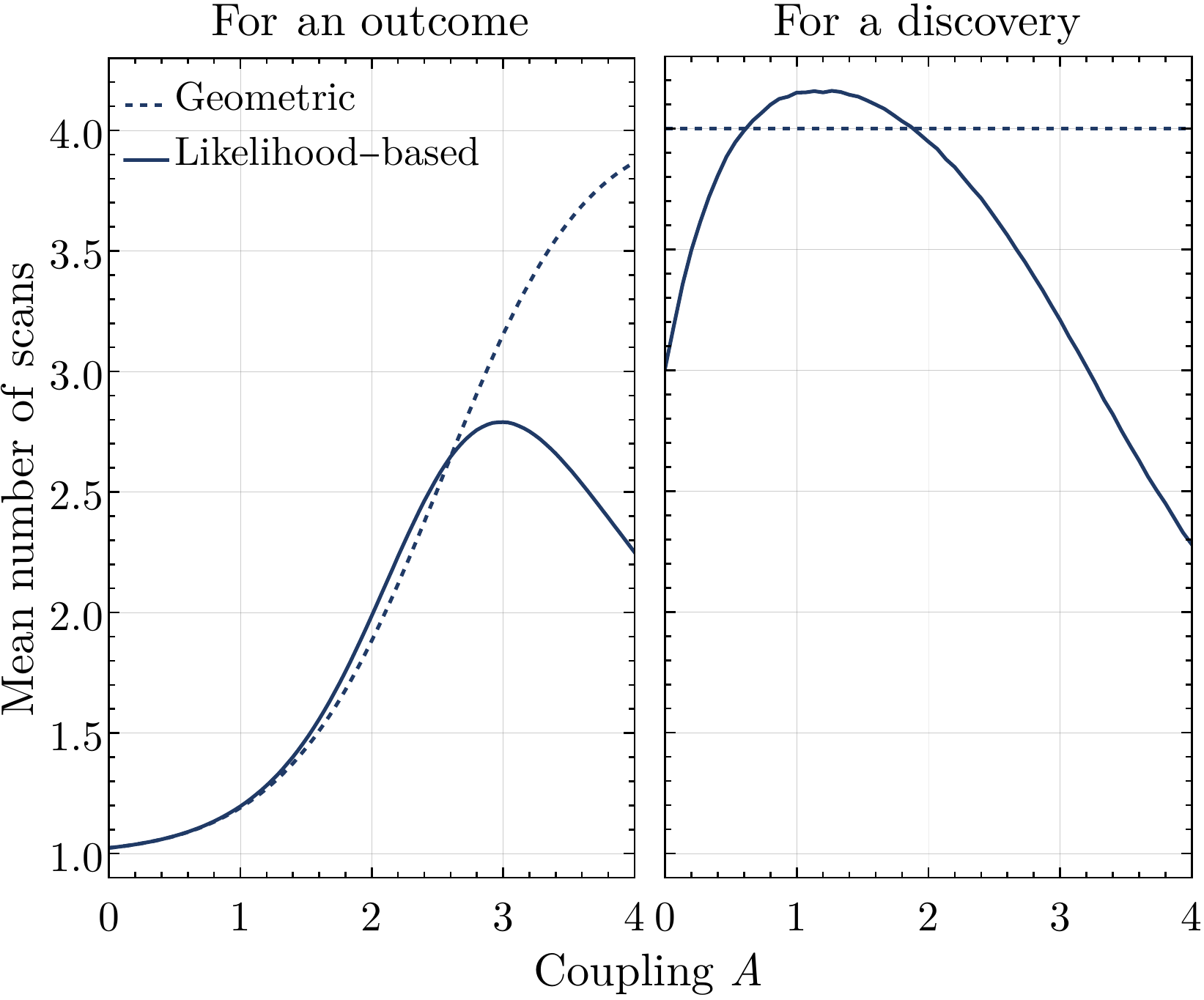}
\caption{Mean number of
scans needed to claim a statistical discovery (right) and 
to get an outcome, either a discovery or a no-discovery claim
(left) for the two analysis approaches (Sections \protect%
\ref{sec:geometric} and \protect\ref{sec:likelihood_rescan}), for different values of the coupling $\Av$.
The setup of the simulation is discussed in Section \ref{sec:comparison}.}
\label{fig:nscan}
\end{figure}

We emphasize once more that this is a toy example used for illustrative purposes but with no realistic
physical meaning. In general, the value of $\Av$ is an alias for
the coupling constant for the interaction channel under
consideration and can carry units (see e.g.\ Section \ref{sec:ulhay}).
For the same reason, the reachable sensitivity depends on
the explicit values of $u_\pel$ \eqref{eq:defUkXk}, which can
be constructed by tailoring the explicit configuration of a given
detector. A fine tuning of the thresholds $x_{\mathrm{thr}}$
is also possible, in order to increase the power for smaller
values of $\Av$. In this sense, analytic expressions
\eqref{eq:powGeo} and \eqref{eq:powLik} can be used to identify optimal choices of these parameters based on the specifics of the experiment being conducted.

Another useful aspect  of the likelihood-based rescan
protocol is that it does not require to perform all the $k_\alpha$ scans in
order to claim a statistical discovery. Hence, it has the potential to reduce
substantially the costs associated with additional rescans.
Specifically, while  a discovery claim via the geometric approach
always requires us to perform $k_\alpha-1$ scans
(see Figure \ref{fig:digG}), on average, we may expect 
the average number of scans required for a statistical discovery to be approximately the same or
lower for the likelihood-based rescan protocol.
This is illustrated on the right panel of Figure
\ref{fig:nscan} by means of a Monte Carlo simulation for our toy example. The left panel,  shows that the average number of scans required to reach a
conclusion (either a discovery or a no-discovery claim) is lowered when relying of the likelihood-based rescan protocol for large values of the coupling.
Whereas, given the initial tuning, it is the same for both
frameworks as $A\to 0$.

\begin{figure*}[t]
\includegraphics[width=\textwidth]{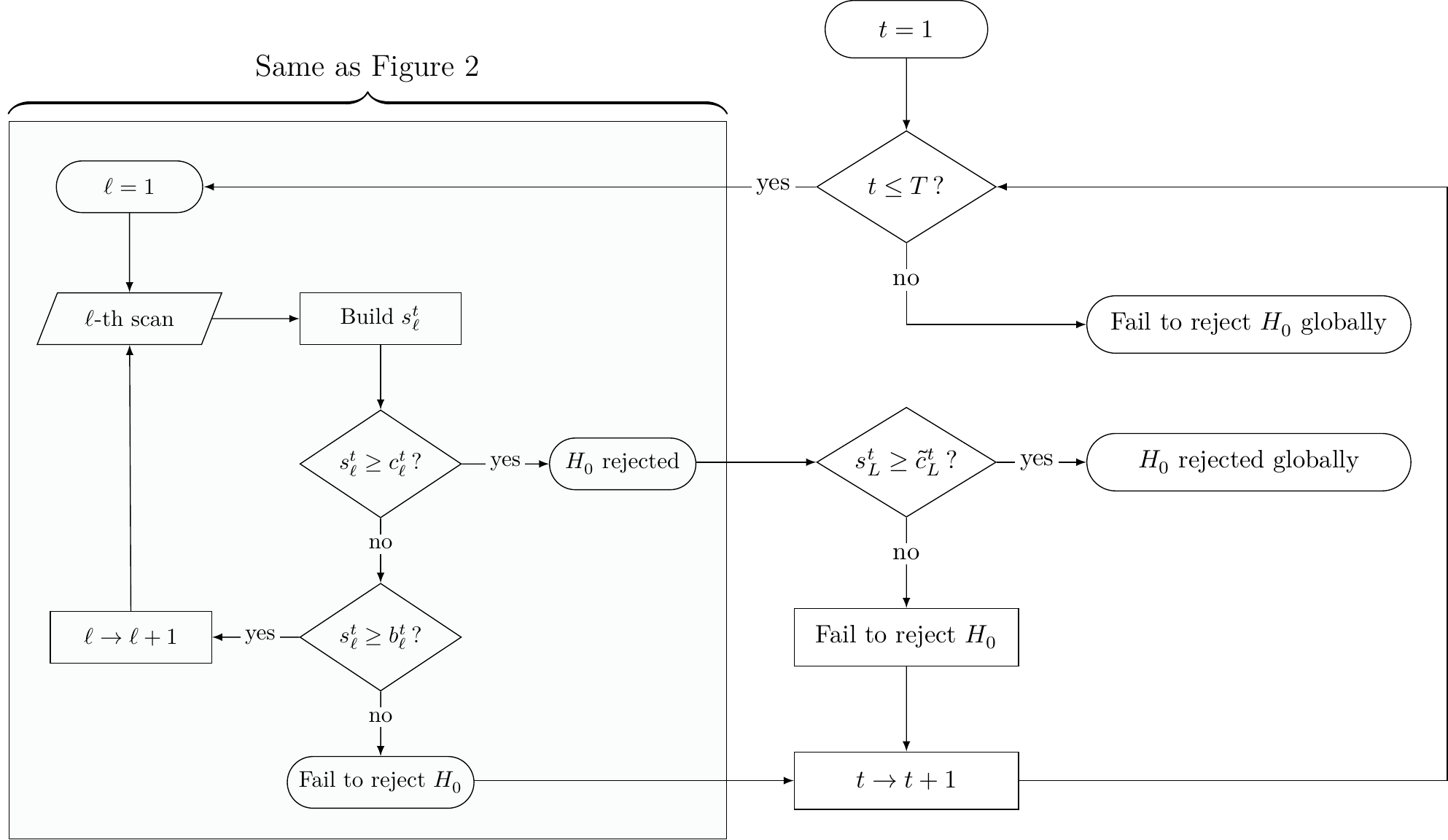}
\caption{Flowchart of the likelihood-based rescan protocol
(Figure \protect\ref{fig:digL}) corrected to account
for the scan of multiple frequencies.}
\label{fig:digLEE}
\end{figure*}

\section{Look-elsewhere effect corrections}
\label{sec:LEE}

The protocols discussed in Section \ref{sec:rescan} can be
meaningfully applied with no further modifications only when the interest is in testing  \emph{only
one} given frequency. In the more realistic scenario where a set of
$T\leq M$ frequencies $\{\nu_{\fre}\}_{\fre=1}^{T}$ is considered,
the significance must be adjusted to account for the so-called
\emph{look-elsewhere effect} (LEE). That is, the probability that
a random fluctuation may cross the discovery threshold anywhere
over the frequency range considered. In other words, one must
ensure that the probability of a false discovery at any
frequency $\nu_1,\,\dots,\,\nu_T$ does not exceed the desired 
significance level $\alpha$.

While several solutions to address the LEE have been proposed in literature
\cite[e.g.,][]{Algeri:2016gtj,Foster:2017hbq, GV10}, they are only
applicable to the situation where just one scan is performed. In what follows, we discuss different approaches that allow us to correct for the  LEE even when the data are collected sequentially over multiple scans.

The simplest possible solution is that of re-adapting classical approaches such as \v{S}id\'ak's and Bonferroni corrections \cite[e.g.,][]{Algeri:2016gtj} to our context. This can be done by replacing $\alpha$ in \eqref{eq:PUDn} with $1-(1-\alpha)^{1/T}$
for \v{S}id\'ak and with $\alpha/T$ for Bonferroni.
The former provides an exact result, but relies on the assumption that all the test being performed 
are independent. Whereas, the latter trades the assumptions of independence for a much more conservative result, and, consequently, lower power.

Unfortunately, in axion searches, the $T$ tested frequencies are
not independent. That is because the location of the axion is
unknown, and thus, the step at which the frequency range is
scanned is usually smaller than the expected width of the signal.
This is done to ensure that no power is shared between two regions
and thus dampened. Nonetheless, one can overcome this limitation by
implementing \v{S}id\'ak or Bonferroni corrections based on the effective number
of independent regions, $R_{T}^\alpha$. The latter can be defined
as the value of independent tests that would determine the correct
significance when replacing $\alpha$ in \eqref{eq:PUDn} with  
$R_{T}^\alpha$ \eqref{eq:PDaN} to compute the corresponding
constants. Therefore, for a given value of $R_{T}^\alpha$, an
adequate LEE correction can be obtained by replacing the
probability of a discovery claim under $H_0$ at the $\pel$-th
scan in \eqref{eq:PDaN} with
\begin{equation}\label{eq:PDaNLEE}
    P(D_{\fre\pel}|\,H_0)=
    1- \Bigl(1- \frac{\alpha}{k_\alpha}\Bigl)^{\sfrac{1}{R_{T}^\alpha }}\approx
    \frac{\alpha}{R_{T}^\alpha k_\alpha}
\end{equation}
in order to have
\begin{equation}\label{eq:PUDnLee}
    \sum_{\fre = 1}^{T}
    \sum_{\pel = 1}^{k_\alpha} P\left(D_{\fre\pel}\middle|\,H_0\right)
    =\alpha
\end{equation}
for each of the frequencies, $\fre=1,\,\dots,\,T$, being tested.
Hereinafter, we will refer to this framework as the
\emph{independent regions} approach.
In the latter, the number of independent regions, $R_{T}^\alpha$,
depends on both the number of scanned frequencies,
$T$, and the significance level, $\alpha$. Therefore,
the procedure requires a Monte Carlo calibration of  $R_{T}^\alpha$ on the basis of 
\eqref{eq:PUDnLee}  --- see e.g.\ \cite{Foster:2017hbq}. 

Here, we propose an alternative solution that allows us to overcome
all the limitations of the above-mentioned approaches. Our strategy
consists of constructing an updated version of the constants
$c_\pel$ that accounts for the LEE, while leaving the
likelihood-rescan protocol (Figure \ref{fig:digL}) at each of
the $T$ frequencies unchanged. This can be done following the 
scheme on Figure \ref{fig:digLEE}; the main steps are
described below. 

For a given frequency, $\fre$, let   $L_\fre$ be the scan at
which an outcome is reached when following the rescan-protocol
in Section \ref{sec:likelihood_rescan} and  Figure \ref{fig:digL}. Denote with $s_{\lst}^\fre$ the value of the test statistic at such scan; to ease the notation,
the subscript $\fre$ has been dropped
from $L_\fre$ 
since the dependency on frequency is
reminded by the superscript.
To control for the look-elsewhere effect we must guarantee that
\begin{equation}\label{eq:PLEE1}
    P\left(\bigcup_{\fre = 1}^T 
    \left\{s_{\lst}^\fre - c_{\lst}^\fre\geq 0\right\}
    \middle| H_0\right)\leq\alpha,
\end{equation}
which is equivalent to the condition
\begin{equation}\label{eq:PLEE2}
    P\left(
    \max_{\fre = 1,\dots,T}
    \left\{s_{\lst}^\fre - c_{\lst}^\fre\right\} > 0
    \middle| H_0\right)\leq\alpha.
\end{equation}

This leads to the definition of the new (global) statistic
\begin{equation}\label{eq:bigS}
    \mathcal{S} = \max_{\fre = 1,\dots,T}
    \left\{s_{\lst}^\fre - c_{\lst}^\fre\right\}.
\end{equation}
which is simply  the maximum of the stochastic process
$\left\{s_{\lst}^\fre - c_{\lst}^\fre\right\}_{\fre=1}^T$. 
Notice that, while each $s_\pel$ in \eqref{eq:sl} is Gaussian,
$s^{\fre}_{\lst}$ is not; that is because its distribution depends on the (binary)
outcome of the decision rule in \eqref{eq:Cl}, and thus the value
$\lst_{\fre}$ at which the decision is reached is itself random.
Unfortunately, the distribution of  $\mathcal{S}$ cannot be
easily derived explicitly, nor it can be  approximated using the
upcrossings/Euler characteristic heuristic typically used in the
context of LEE corrections \cite{GV10,Algeri:2019}.
Nonetheless, it is possible to retrieve the null distribution of
$\mathcal{S}$ by means of a Monte Carlo simulation. 

Specifically, we are interested in estimating the quantile
of order $1 - \alpha$ of $\mathcal{S}$, i.e.\ $\mathcal{S}_{1 -\alpha}$ for
which \[P(\mathcal{S}>\mathcal{S}_{1 - \alpha})=\alpha.\]
We can then construct our newly ``LEE-updated'' constants,
namely $ \tilde{c}_{\lst}^\fre$, as
\begin{equation}
\label{eqn:new_thresholds}
    \tilde{c}_{\lst}^\fre = c_{\lst}^\fre+\mathcal{S}_{1-\alpha}.
\end{equation}
Finally, a statistical discovery at (global) $\alpha$ significance is claimed
if $s_{\lst}^\fre\geq \tilde{c}_{\lst}^\fre$ for at least one
frequency $t=1,\,\dots,\,T$. The main steps of this approach are
summarized in Figure \ref{fig:digLEE}.

\begin{figure}[b!]
\includegraphics[width=\columnwidth]{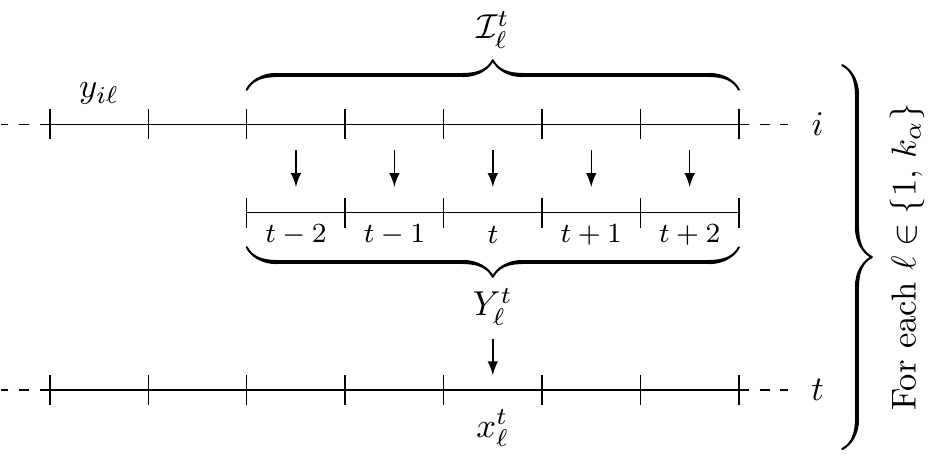}
\caption{Scheme of the setup assumed in the toy
example of Section \protect\ref{sec:lee_apppli}.}
\label{fig:seq1}
\end{figure}

\subsection{Statistical properties of different LEE corrections}
\label{sec:lee_apppli}

In this section, we investigate the properties of the look-elsewhere
corrections described above by extending the toy example introduced
in Section \ref{sec:comparison} to allow an analysis over multiple
frequencies. Specifically, we start
once again by considering a $5\sigma$
significance level ($\alpha=2.87\times 10^{-7}$) and $k_\alpha = 5$,
so that the  constants $b_\pel$ and $c_\pel$ are those
reported in Table \ref{tab:table1}. The setup considered is
illustrated in Figure \ref{fig:seq1}, and is intended to emulate as closely as possible a real axion experiment.

The data consist of a string of noise fluctuations $y_{i\pel}$,
with the index $i$ marking the frequency $\nu_i$ at which the datum
is taken; the subscript $\pel=1,\,\dots,\,k_\alpha$ indexes the rescan. For simplicity, we consider only one detector configuration, i.e., $N_i=1$ for all
$i=1,\,\dots,\,M$, and thus the subscript $j$ is dropped. 

We  assume that $y_i\sim\mathcal{N}(\mu_i=0,\sigma_i=1)$, under $H_0$. Whereas,  the
alternative hypothesis, $H_1$, corresponds to a signal $\mu_i = \Am$,
spanning over the  neighborhood
$\mathcal{I}_{\pel}^{\fre}$ of $\nu_\fre$. The latter consists of 
a row of $d = 5$ frequency bins centered around $\nu_\fre$. For each frequency $\nu_\fre$,
$\fre=1,\,\dots,\,T$, being tested, the respective set of fluctuations  $Y_{\pel}^{\fre}$ is collected over  $\mathcal{I}_{\pel}^{\fre}$.
The quantities
$x_{\pel}^{\fre}=\sum_{i=t-2}^{t+2} y_{i\pel}$
are then computed; whereas, the variances $u_{\pel}^{\fre}$ are constant and
equal to $1/5$ for each $\fre=1,\dots,T$ and $\pel=1,\dots,k_\alpha$.
Once an outcome is reached, $\nu_\fre$ is shifted by one bin
and the procedure is repeated.
Notice, once again, that $Y_{\pel}^{\fre}\cap Y_{\pel}^{\fre+1}
\neq\varnothing$, thus the corresponding  $x_{\pel}^{\fre}$
and $x_{\pel}^{\fre+1}$ (and the respective tests) are not independent.

\begin{figure}[t]
\includegraphics[width=0.95\columnwidth]{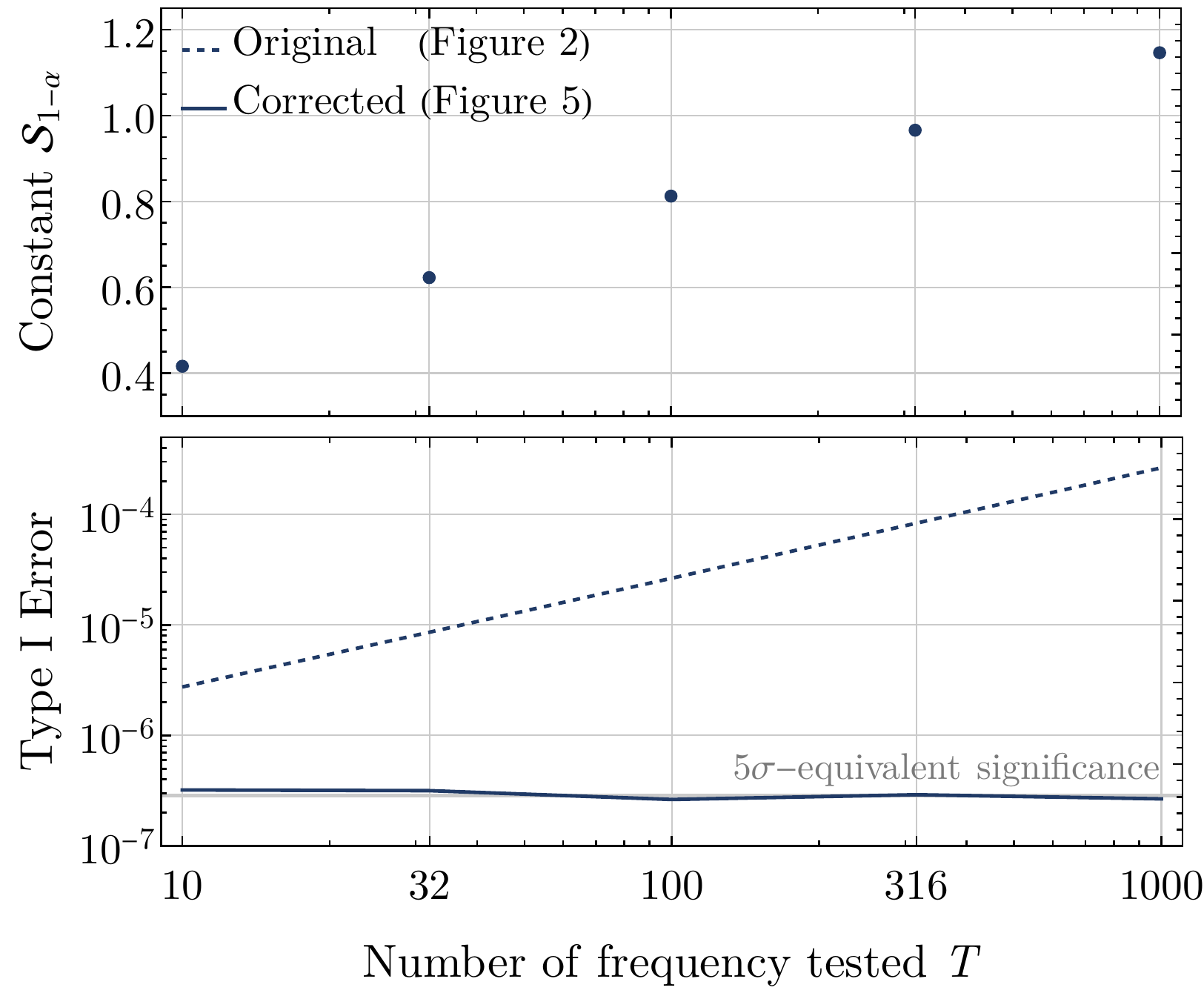}
\caption{Top: Quantile of order $1-\alpha$ of the test statistic
$\mathcal{S}$ in \protect\eqref{eq:bigS}, as a function of the different
number of frequencies $T$ tested. Bottom: Probabilities of
false discoveries for the proposed LEE corrections based on
\eqref{eqn:new_thresholds} (solid line) and those obtained when
conducting only a localized analyses at each individual frequency
with no LEE correction (dashed line).}
\label{fig:lete}
\end{figure}

We implement the LEE corrections based on \eqref{eqn:new_thresholds} where the quantile $\mathcal{S}_{1-\alpha}$ is estimated by means of a Monte Carlo simulation. The results obtained considering an increasing number of total frequencies, $T$, being tested  are shown in the upper panel
of Figure \ref{fig:lete}. 

The bottom panel of Figure \ref{fig:lete} shows that
the LEE corrections based on \eqref{eqn:new_thresholds} (solid line)
ensure that the probability of false discovery is equal to the predetermined $5\sigma$ significance
level.  The same is not true when ignoring the LEE and conducting a local  analysis at each individual frequency (dashed line). In this case, the probability of a false discovery  increases with the number of tested frequencies.  

\begin{figure}[t]
\includegraphics[width=\columnwidth]{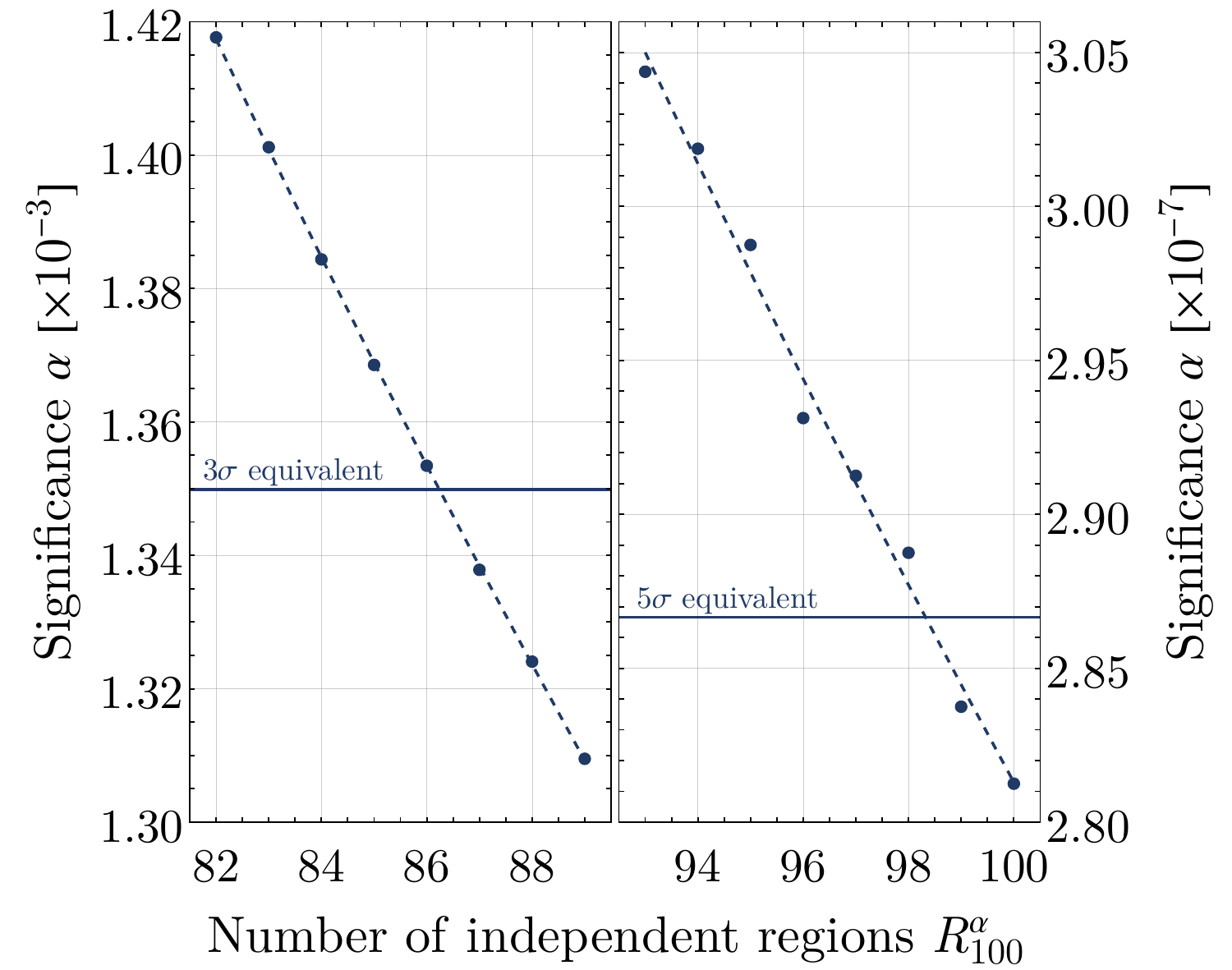}
\caption{Probability of false discovery evaluated via Monte Carlo, when testing
$T=100$ frequencies at $3\sigma$ equivalent significance (left)
and $5\sigma$ (right), as a function of independent regions
$R_{100}^{\alpha}$ considered.}
\label{fig:mreg}
\end{figure}

For the sake of comparison, we also implement the LEE corrections based on  the
number of independent regions $R_{T}^\alpha$ (see  Equation \eqref{eq:PDaNLEE}). In this case,  it is necessary to ``tune'' $R_{T}^\alpha$ by means of a Monte Carlo simulation.
Figure \ref{fig:mreg}, shows the probability of false discovery obtained when testing $T=100$ frequencies. Recall that, in this case, the analytical framework is the same as that in Figure \ref{fig:digL} with the  
constants $c_\pel$ and $b_\pel$ computed by assuming
different values of $R_{T}^\alpha$ in order to ensure the validity of \eqref{eq:PDaNLEE}. On the left panel of Figure \ref{fig:mreg}, we consider a
 significance level of $3\sigma$, while the right panel corresponds to $5\sigma$. When choosing $R_{T}^\alpha=T=100$
the result is conservative (sensitivity loss); whereas, as
$R_{T}^\alpha$ decreases the more we are susceptible to
noise-triggered false claim. A linear fit applied to the
Monte Carlo output shows that the   number of independent region  is
$R_{100}^{5\sigma} \approx 98.3$ for a $5\sigma$ significance
while it is sensibly lower in the $3\sigma$ case:
$R_{100}^{3\sigma} \approx 86.2$.

\begin{figure}[t]
~\\[0.85em]
\includegraphics[width=\columnwidth]{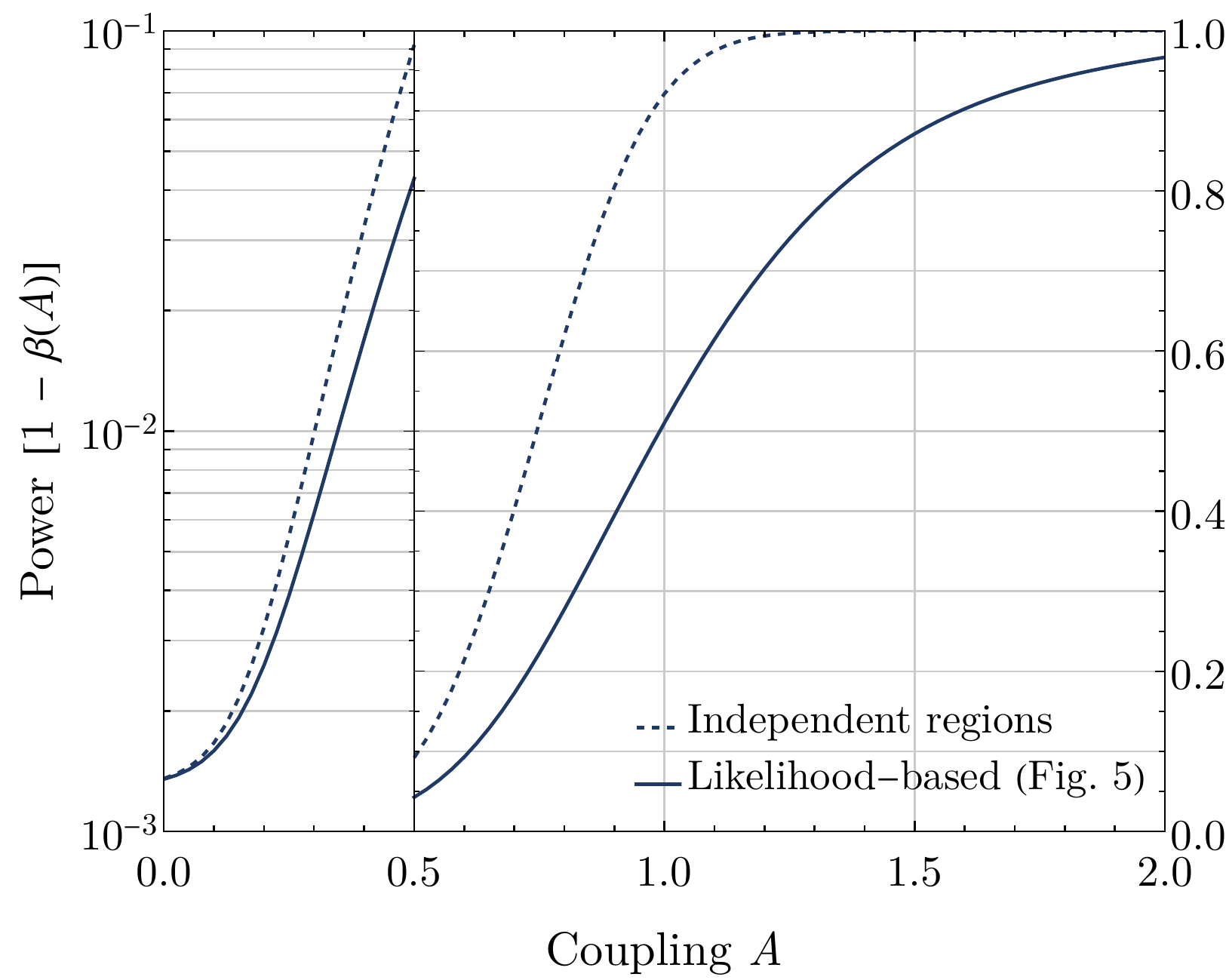}
\caption{Power as a function of the coupling $\Av$ for a
significance of $3\sigma$. Solid line: Corrected
likelihood-based approach (Figure \protect\ref{fig:digLEE}).
Dashed line: likelihood-based approach (Figure
\protect\ref{fig:digL}) with $R_{100}^{3\sigma} \approx 86.2$
independent regions --- see \protect\eqref{eq:PDaNLEE}. The values
in the range $\Av\in[0,\,0.5]$ are magnified in logarithmic scale.}
\label{fig:powLee}
\end{figure}

Figure
\ref{fig:powLee} compares the power of the LEE corrections based on \eqref{eqn:new_thresholds} and those based on the independent regions approach  at $3\sigma$ significance. Among the two, the  independent region approach clearly enjoys  higher power.
This may be due to different reasons. Firstly, the procedure
based on the independent regions lacks of what we may call
``dangling outcomes''. That is, in the LEE framework of Figure
\ref{fig:digLEE}, we may encounter the situation where
\begin{equation}\label{eq:dangling}
    c_\lst^\fre\leq s_\lst^\fre<\tilde{c}_\lst^\fre.
\end{equation}
In this case, the procedure
stops and no discovery is claimed, even if the tested frequency
has not been ruled out as a promising candidate. In fact, if
only $s_\lst^\fre$ had fluctuated below $c_\lst^\fre$ (and
$\lst<k_\alpha$) further scans would have been performed.
Clearly, we may expect condition \eqref{eq:dangling} to
be  in the ``gray zone'' between a very weak and very strong coupling. If $\Av\to 0$, we expect
$s_\ell^\fre < b_\ell^\fre < c_\ell^\fre$ most of the times; whereas,  if $\Av\to\infty$,   we  expect
$s_\lst^\fre\gg\tilde{c}_\lst^\fre$.

\begin{figure}[t]
\includegraphics[width=0.945\columnwidth]{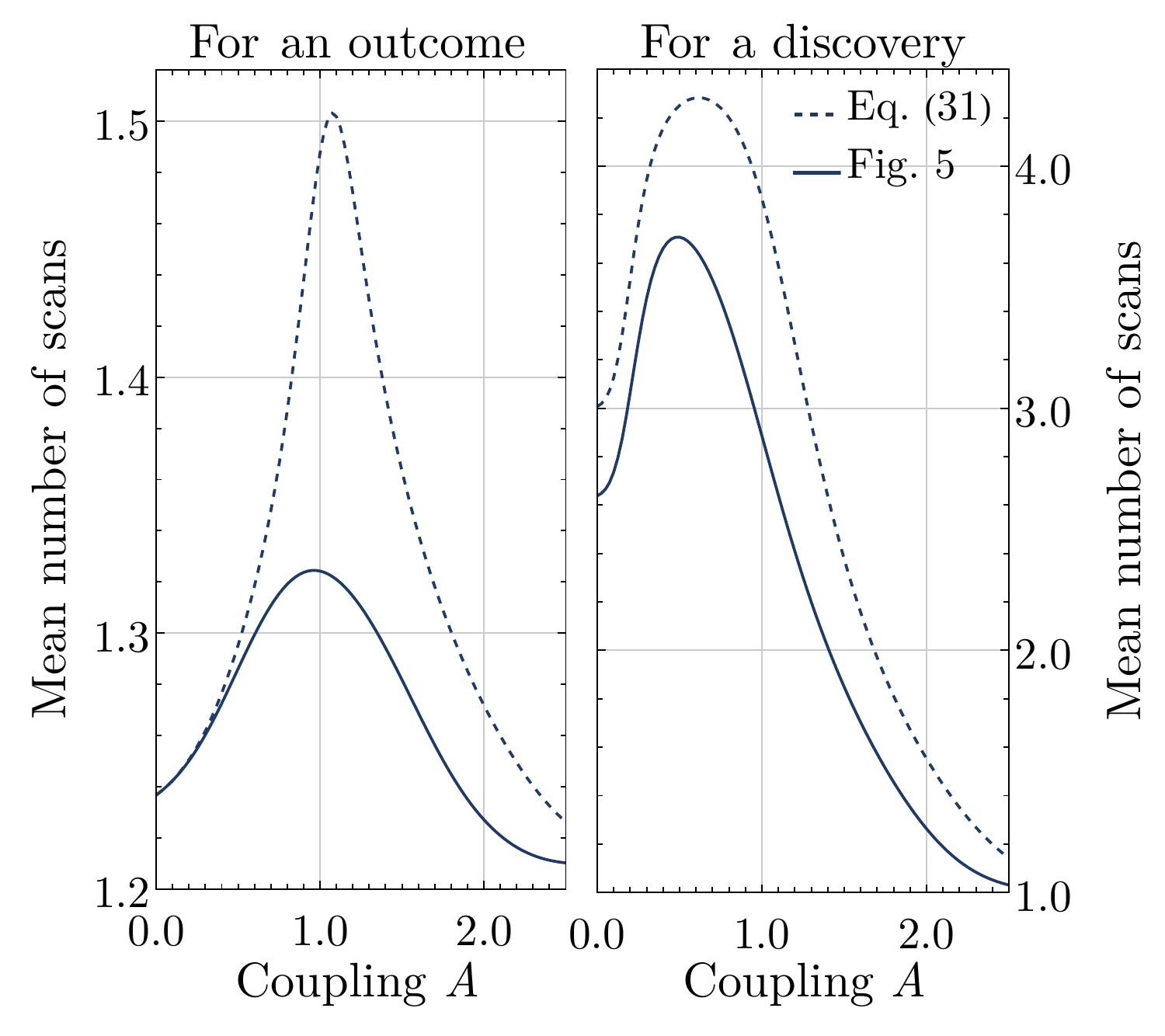}
\caption{Mean number of
scans needed to claim a statistical discovery (right) and 
to get an outcome, either a discovery or a no-discovery claim
(left) for the approach outlined in Figure \protect%
\ref{fig:digLEE} (solid) and the one based on the
application of the independent regions (dashed), for different values of the coupling $A$.}
\label{fig:nsclee}
\end{figure}

As highlighted in Figure \ref{fig:nsclee}, the higher power
reached by the independent-region procedure is also due to
its tendency to perform more
scans. In terms of number of scans needed, a peak is achieved in the transition
between low to high powers (see e.g.\ left panel of Figure
\ref{fig:nsclee}). The maximum difference in the
number of scans per outcome occurs at approximately  $\Av=1$ and
corresponds to the highest relative difference between
the two powers. In other words, the power increases with
the number of performed scans, which is not surprising.

Everything highlighted so far applies to the case of a toy example.
In a real experiment, the inferential procedure adopted needs to be tailored
on the specific setup to which it is applied. Moreover,  the performance of different approaches 
is bounded by
practical constraints such as measurement time, sensitivity reach,
or tuning. A direct comparison among them is hard to establish,
since in a real setup not all scans are equal  (like the $y_{i\pel}$
in the toy example) and the
precision of each the measurement grows with the time invested in collecting the data. For instance, a procedure that is less sensitive but, on average, requires less scans to reach a conclusion (discovery/no discovery) allows researchers to redistribute the
``extra time'' to make fewer, but more precise, measurements. On this note, a procedure that reduces, at least on average, the number of scans needed is expected to be particularly impactful in view of future axion
detectors, such as ALPHA \cite{Millar:2022peq}, where the projected number
of frequencies to be tested is on the order of $10^7$.

Finally, the CPU time required by each procedure is also non-negligible in practical applications. In this respect, calibrating the number of independent-regions, $R_{T}^\alpha$, via Monte Carlo, requires approximately one order of magnitude more of CPU time than estimating the quantile $\mathcal{S}_{1-\alpha}$ in \eqref{eqn:new_thresholds}. For instance, in our example, computing each point in Figure 
\ref{fig:mreg} takes the same amount of time needed to estimate $\mathcal{S}_{1-\alpha}$. Although not relevant in terms of statistical properties,   the computational complexity may become a decisive factor  when dealing with real data. 

For all the above-mentioned
reasons, what represents ``the best strategy'' strictly depends on the experimental conditions as well as on the time and computational resources available. In light of this, the development of a solution which is optimal with respect to both the number of rescans and the CPU time, while adjusting for the LEE, is left for future work. 

\section{Application on upper limits}
\label{sec:ulhay}

A straightforward application of the proposed framework
to a real detector is when setting upper limits, i.e.,\ 
the case when no significant signal is detected and thus,
the outcome of the experiment is the exclusion of the
couplings that would have likely generated a signal.
This application is particularly useful when dealing when
projecting the reach of future experiments, for which no
data has yet to be produced.
To ease the intuition, let's consider once again the one scan-only
framework of Section \ref{sec:rescan}. To account for the possibility that a signal with some coupling $\Av$ is present, 
the test statistic in \eqref{eq:defSk} can be reformulated as
\begin{equation}
\label{eqn:sAul}
    s(\Av) = \sqrt{u}\left(\frac{x}{u}-\Av\right)
\end{equation}
and is distributed as a standard normal random variable.
For a given observed value, $\hat{s}(\Av)$, of \eqref{eqn:sAul}, the upper limit for the coupling, $\Av\ul$, is the value of $A$ that solves 
\begin{equation}\label{eq:defUL}
   \Phi\bigl(\hat{s}(\Av)\bigl)
    = \alpha\ul,
\end{equation}
with $\alpha\ul$ being desired significance at which we aim to set our upper limit (e.g., $\alpha\ul=0.1$ for a $90\%$ upper limit).
Since $\Phi(\cdot)$ is the distribution of a standard Gaussian,
Equation \eqref{eq:defUL} can be easily inverted. The resulting upper limit for $A$ is then
\begin{equation}\label{eq:defAul}
    \Av\ul = \frac{\hat{x}/\sqrt{u}-\Phi^{-1}(\alpha\ul)}{\sqrt{u}}.
\end{equation}
This result is similar to what already found in 
\cite{Foster:2017hbq}. Equation \eqref{eq:defAul} can be further simplified
 when dealing with projected values of $\hat{x}$ in view of future experiments rather than actual measured data.
In this case, we expect to measure a null power excess ($\hat{x}=0$).
This gives an analytical formula which is statistically justified
and can be easily applied --- see e.g.\ \cite{Millar:2022peq}.

In the remaining  of this section, we demonstrate the
validity of Equation \eqref{eq:defAul} by comparing
the resulting upper limit with the outcome of 
an analysis based on actual data.
We will refer to the first run of HAYSTAC
detector as presented in Refs.\ 
\cite{Brubaker:2016ktl,Brubaker:2017ohw}.
Although the detector has been updated and newer results have
been produced since then --- see e.g.\ \cite{HAYSTAC:2020kwv}
--- Ref.\ \cite{Brubaker:2017ohw} is sufficiently detailed, in terms of
 general properties and data acquisition,   to
produce meaningful results.

\begin{figure}[t]
\includegraphics[width=0.96\columnwidth]{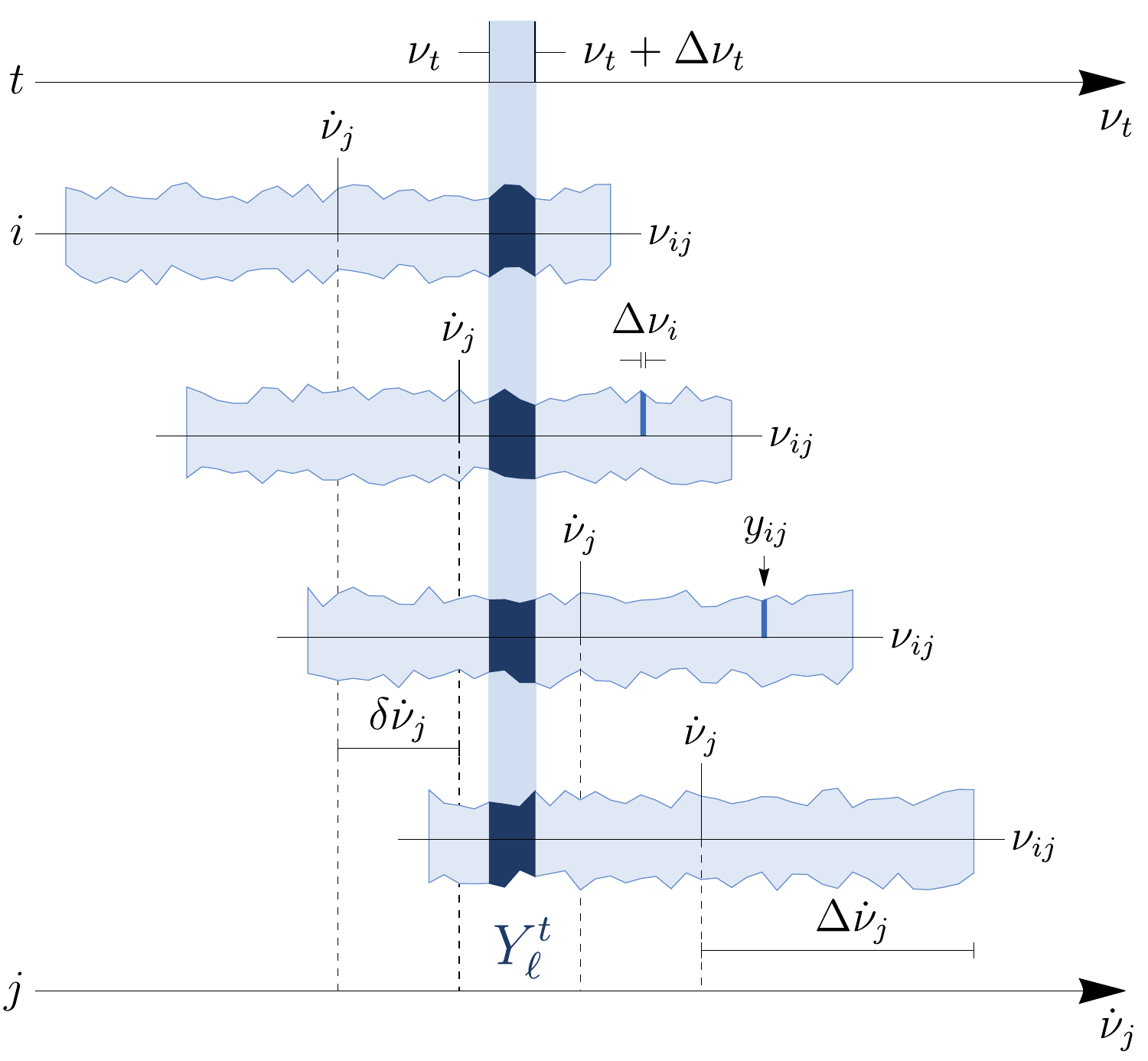}
\caption{Sketch of the spectra, data format and relevant
quantities for a HAYSTAC-like experiment.}
\label{fig:nstack}
\end{figure}

Given its specifics, the experiment is designed to
probe the frequency range
$[5.7,\,5.8]\,\si{\giga\hertz}$. The search is performed
sequentially tuning the resonance of the cavity
$\dot{\nu}_j$ at steps $\delta\dot{\nu}_j =\SI{75}{\kilo\hertz}$.
Following our notation, the subscript $j$ indexes
the resonance of   a given detector configuration and
characterized by different values of its relevant parameters;
for instance, the loaded quality factor $Q_j$ (see also Figure
5.13 of \cite{Brubaker:2017ohw}).

At each configuration, the detector is  sensitive to the range
of frequencies $\nu_{ij}\in [\dot{\nu}_j-\dot{\nu}_j/Q_j,\,
\dot{\nu}_j + \dot{\nu}_j/Q_j]$; the latter correspond to  the overlapping spectra in Figure \ref{fig:nstack}.
We denote with $\nu_{ij}$ the $i$-th frequency probed
in the $j$-th configuration.
Once a resonance is set, the power coming from
a linear amplifier is measured for a total time
$\tau_j=\SI{15}{\min}$, and is averaged over the
time $\tau_{ij}$ needed to reach the frequency
resolution $\Delta\nu_{ij}=1/\tau_{ij}=\SI{100}{\hertz}$.
In the real data acquisition, the whole range is scanned twice,
``followed by several shorter scans to compensate for
nonuniform tuning'' (see \cite{Brubaker:2017ohw}, Figure 6.7).
In order to collect a reasonable amount of data for the
following calculations, we rely on
Section 4.3.5 of the same reference and assume that
the frequency range has been scanned   exactly
$3$ times. In other words, each spectrum in Figure
\ref{fig:nstack} comes in triplets or, conversely, we can
imagine to deal with spectra acquired for 
$\tau_j=\SI{45}{\min}$.

The averaged measured powers in each spectrum can be
shifted and rescaled with respect to their mean baseline,
leading to adimensional quantities (see for instance \ Figure 2(c) in \cite{Brubaker:2017rna}). The resulting
power fluctuations correspond to our  $y_{ij}$ in Equation \eqref{eq:ypdf}
and constitute the data sample used in our analysis (without any further
scaling and/or rebinning).
The standard deviation of each fluctuation $y_{ij}$ is
\begin{equation}\label{eq:defsigma}
	\sigma_{ij}
	= \sqrt\frac{\tau_{ij}}{\tau_j}
	= \frac{1}{\sqrt{\tau_j\Delta\nu_{ij}}}.
\end{equation}
Whereas, their mean expresses the signal-to-noise
ratio and can be factorized as \eqref{eq:muij}.
Here, $\Av$ is an alias of the physical coupling constant
between an axion and two photons, i.e.\ $\Av=g_{a\gamma\gamma}^2$.
Note that, in natural units, $g_{a\gamma\gamma}$ is usually
expressed in \si{\per\electronvolt}. 
The factor $w_{ij}^{\fre}$ is nonzero only if the
axion at  frequency $\fre$  deposits power on the
$i$-th bin of the $j$-th spectrum. Explicitly:
\begin{equation}\label{eq:defwh}
	w_{ij}^{\fre} =
	\frac{P_j(\nu_{\fre})\,\sqrt{\tau_{ij}/\Delta\nu_{ij}}}
	{h\dot\nu_{j}\,N_\mathrm{sys}(\dot\nu_j, \nu_{ij})}
	\,D(\nu_{\fre},\,\dot\nu_j)\,L(\nu_{\fre},\,\nu_{ij}).
\end{equation}
The parameterization of each term follows from
Ref.\ \cite{Brubaker:2017ohw}. Specifically, $P_j$ is the power per unit
coupling constant (4.31), $N_{\mathrm{sys}}$ are the noise
quanta (Figure 6.9 in \cite{Brubaker:2017ohw}), $D(\nu_{\fre},\,\dot\nu_j)$ as in Equation (5.2) of \cite{Brubaker:2017ohw}
accounts for the possible mismatch between $\dot\nu_j$ and
$\nu_{\fre}$, while $L(\nu_{\fre},\,\nu_{ij})$ is the integral of
the signal lineshape as in Equation (7.13) of \cite{Brubaker:2017ohw} between each bin's edges. That is,
since $\Delta\nu_{\fre} = 10^{-6}\nu_{\fre}\gg\Delta\nu_{ij}$,
it quantifies the amount of signal falling in the bin.

\begin{figure}[t]
\includegraphics[width=0.96\columnwidth]{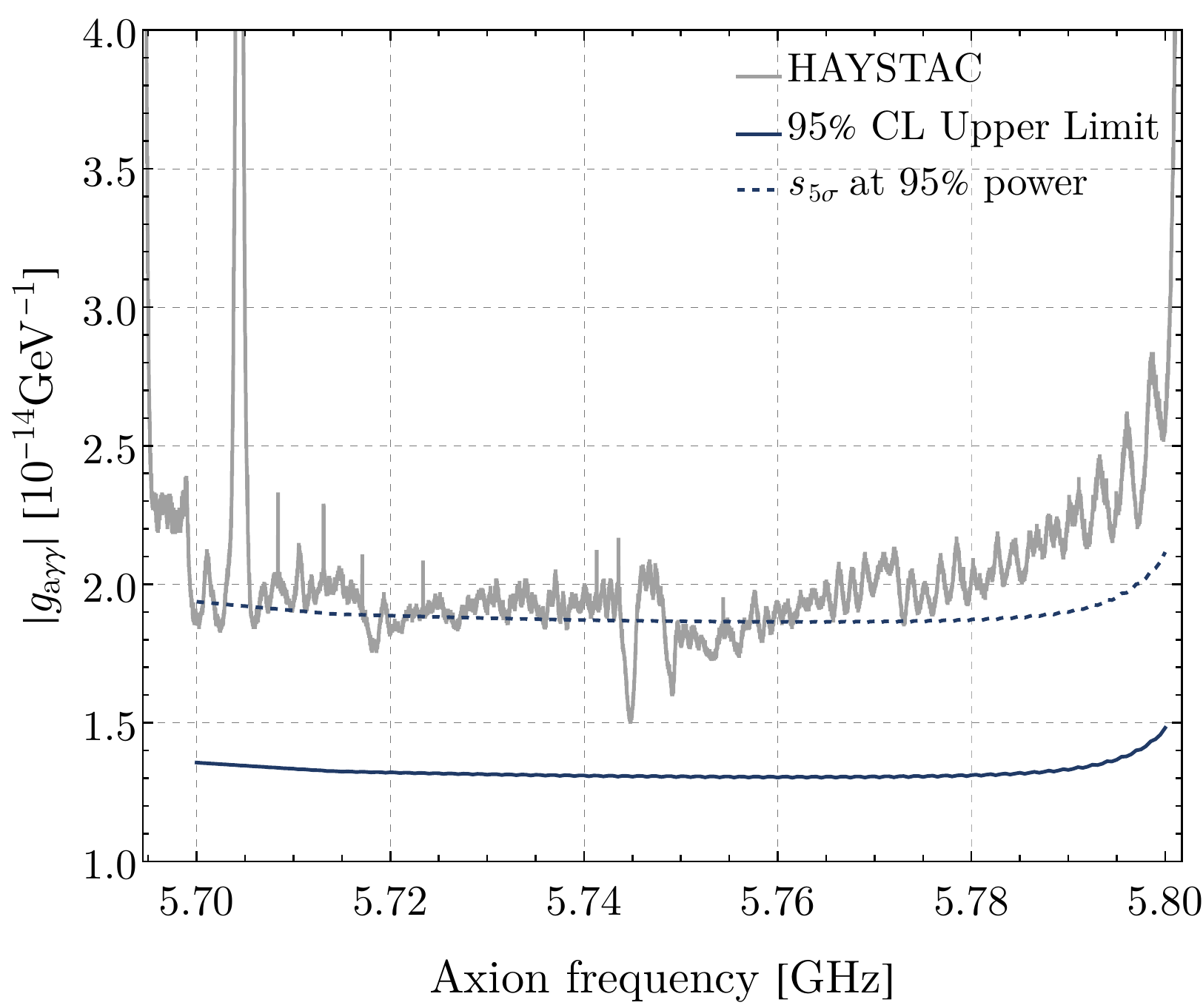}
\caption{Upper limit projections for the first run of HAYSTAC
detector originally presented in Refs.\ \protect%
\cite{Brubaker:2016ktl,Brubaker:2017ohw} (gray line).
Blue solid line line: 95\% upper limit from Equation
\protect\eqref{eq:defAul} with $\hat{x}=0$. Dashed line:
limit given by $s_{5\sigma} = 5$ at 95\% power.}
\label{fig:haystack}
\end{figure}

Once the frequency $\nu_{\fre}$ to be tested has been selected, we can  calculate the quantity $u^{\fre}$ in \eqref{eq:defUkXk}
by collecting the measurements obtained over the bins in the spectrum that falls within the
range $\nu_{\fre}\leq\nu_{ij}\leq\nu_{\fre}+\Delta\nu_{\fre}$.
We can then compute Equation
\eqref{eq:defAul} with $\hat{x}=0$ to obtain the projection of the
95\% upper limit. The latter corresponds to the solid blue line in Figure \ref{fig:haystack}. On the other hand, if we chose to define
the upper limit with respect to the threshold value for a
statistical discovery, $s_{5\sigma} = 5$, we would end up with a
result about a factor 1.45 higher, as shown  in Figure
\ref{fig:haystack} as  dashed blue line. The definition of the
upper limit in terms of a threshold value is the same as that adopted
in Ref.\ \cite{Brubaker:2017ohw}. Their result is
also shown in Figure \ref{fig:haystack} as a gray curve.
The results are quite compatible, even if our computation
relies on many simplifications and parameterizations of
time-dependent value describing the detector properties.

\section{Conclusions}
\label{sec:conclusion}

In this paper, we have presented an inferential framework for
axion searches that is statistically valid even when the experimental setup involves repeated scans. In Section
\ref{sec:likelihood}, we have introduced the test statistic which
is at the core of the likelihood-based rescan protocol discussed
in Section \ref{sec:likelihood_rescan} and is summarized in Figure
\ref{fig:digL}. Specifically, when several scans  are performed sequentially, our test statistic and the corresponding discovery thresholds are updated  with as new sets of data are collected, until the fluctuation is
reabsorbed or a statistical discovery is claimed. An important feature of the proposed procedure is that, when the analysis is performed at a fixed frequency, one can compute the power and the statistical significance analytically. Adequate comparisons with the so-called geometric approach  have been conducted  in Section \ref{sec:comparison} and have shown that the 
the likelihood-based
inference enjoys higher power and, on average,  a lower number of scans. Moreover,
the likelihood-based inference provides a higher freedom in
the definition of relevant quantities for data-acquisition. For
instance, when the significance $\alpha$ and the maximum number
of scans $k_\alpha$ have been set, the threshold $x_{\mathrm{thr}}$ in 
\eqref{eq:condition2} is automatically determined and so is the rescan
probability $p_0$. On the other hand, the likelihood-based approach
allows for more freedom in specifying $p_0$ given
$\alpha$ and $k_\alpha$ fixed, as one can tune the constants $b_\pel$ and $c_\pel$ in Equations \eqref{eq:FindCn}--\eqref{eq:FindBn} to achieve the desired significance.

Another important contribution of this work is the introduction of adequate corrections for look-elsewhere
effect, and required when testing multiple axion
frequencies $\nu_\fre$. Specifically, in Section \ref{sec:LEE}, we have  discussed   Bonferroni and Sidak's
correction based on the effective number of independent
regions, and we have introduced a novel alternative solution that is meant to ease the
burden of the Monte Carlo tuning. The two approaches have been compared  in Section \ref{sec:lee_apppli} by means of a toy example. The latter has shown that, while the
method based on the independent regions is characterized
by a higher power, it is also requires, on average,  a higher  number
of scans per outcome and discovery. It has to be noted, however, that establishing
a direct comparison between the two in a real setup is a non-trivial task due to the  interplay between the number of tested frequencies, the
time spent measuring and the reachable precision of each
measurement. An in-depth study of the performance
of the different approaches when applied to  real experiments is of paramount importance and paves the way to future
improvements of this work.

Finally, in Section \ref{sec:ulhay}, we used our frequentist
likelihood-based framework to reproduce a raw projection of
the upper limit reachable by a realistic axion run in a
real detector, as the phase one of HAYSTAC.

\section*{Acknowledgments}
AGR and JC have been supporten by a grant of the Knut and Alice Wallenberg Foundation and the Swedish Research Council.

\printbibliography

\end{document}